\documentclass[usenatbib]{mn2e_letter}
\usepackage{amssymb, amsmath}
\usepackage{graphicx}

\newcommand{\dif}{\mathrm{d}}

\def\la{\langle}
\def\ra{\rangle}

\def\Mo               {\hbox{$M_{\odot}$}}

\def\power{\textrm{erg}\;\textrm{cm}^{-3}\,\textrm{s}^{-1}}

\def\bdotr{\langle|\boldsymbol{\hat{b}}\cdot\boldsymbol{\hat{r}}|\rangle}
\newcommand       \be           {\begin{equation}}
\newcommand       \ee           {\end{equation}}

\newcommand{\acknowledgments}{\begin{small}
    \section*{Acknowledgments}\end{small}}
\newcommand\altaffilmark[1]{$^{#1}$}
\newcommand\altaffiltext[1]{$^{#1}$}

\begin{document}
\title[Turbulent Pressure Support in Galaxy Clusters]{Turbulent Pressure Support in the Outer Parts of Galaxy Clusters}
\author[Parrish, McCourt, Quataert, and Sharma]{
\parbox[t]{\textwidth}{
Ian J. Parrish\altaffilmark{1}\thanks{E-mail: iparrish@astro.berkeley.edu}, Michael McCourt, Eliot Quataert, and Prateek Sharma\altaffilmark{2}}
\vspace*{6pt}\\
\altaffiltext{1}{Department of Astronomy and Theoretical Astrophysics
  Center, University of California
  Berkeley, Berkeley, CA 94720} \\
\altaffiltext{2}{Present Address: Department of Physics, Indian Institute of Science, Bangalore 560012, India} \\
}
\maketitle
\begin{abstract}
  We use three-dimensional MHD simulations with anisotropic thermal
  conduction to study turbulence due to the magnetothermal instability
  (MTI) in the intracluster medium (ICM) of galaxy clusters. The MTI
  grows on timescales of $\lesssim 1$ Gyr and is capable of driving
  vigorous, sustained turbulence in the outer parts of galaxy clusters
  if the temperature gradient is maintained in spite of the rapid
  thermal conduction.  If this is the case, turbulence due to
  the MTI can provide up to 5--30\% of the pressure support beyond
  $r_{500}$ in galaxy clusters, an effect that is strongest for hot, massive clusters.  The turbulence driven by the MTI is
  generally additive to other sources of turbulence in the ICM, such
  as that produced by structure formation.  This new source of
  non-thermal pressure support reduces the observed Sunyaev-Zel'dovich
  (SZ) signal and X-ray pressure gradient for a given cluster mass and
  introduces a cluster mass and temperature gradient-dependent bias in SZ and X-ray
  mass estimates of clusters.  This additional physics may also need
  to be taken into account when estimating the matter power spectrum
  normalization, $\sigma_8$, through simulation templates from the
  observed amplitude of the SZ power spectrum.
\end{abstract}
\section{Introduction}\label{sec:intro}
Clusters of galaxies are on the exponential tail of the mass function
of gravitationally bound objects in the universe.  As a result,
fundamental cosmological parameters, e.g., $\sigma_8$, are very
sensitive to the statistics and evolution of the cluster
population. 
Unfortunately, only $\sim \!\!12\%$ of the total mass of a cluster is
contained in the hot intracluster medium (ICM) that is readily
observable.  Thus for many observational probes the total cluster mass
must be estimated by either fitting to hydrostatic equilibrium or by
assuming scaling relations between total mass $M$ and other
observables, e.g., $M$--$L_x$, where $L_x$ is the total X-ray
luminosity.

A new window for studying galaxy clusters has recently been opened
with measurements of the Sunyaev-Zel'dovich (SZ) effect, which is
produced by the inverse Compton scattering of CMB photons off the hot
ICM.  Recently a variety of surveys, including the South Pole
Telescope (SPT) and the Atacama Cosmology Telescope (ACT) have
compiled the first SZ-selected cluster samples (e.g.,
\citealt{vanderlinde10}).  The SZ signal scales linearly with gas
density (proportional to total pressure) allowing observations to
explore larger cluster radii than is typically possible with X-ray
observations.  A disadvantage of moving to larger radii is that one is
potentially more susceptible to the turbulence and bulk flows that
result from structure formation processes.  These sources of
non-thermal pressure support may significantly complicate the
extraction of cosmological parameters from SZ surveys \citep{shaw10}.

We have learned in recent years that convection in the ICM is very
different from the more familiar convection in stars.  In a dilute,
magnetized medium like the ICM, the mean free path is much larger than
the gyroradius and therefore thermal conduction is entirely
anisotropic along magnetic field lines.  In this regime the
magnetothermal instability \citep[MTI;][]{bal00, ps05} can tap into
the radially decreasing temperature gradient to drive convection
regardless of the background entropy gradient.  The MTI in Cartesian
simulations with a fixed temperature gradient is capable of driving
vigorous convection (approaching supersonic velocities) and a magnetic
dynamo that is similar to that of adiabatic convection \citep{mpsq11}.

\citet{psl08} previously studied the MTI in global galaxy cluster
models.  In their simulations, the rapid thermal conduction made the
outer parts of the cluster isothermal after $\sim \!1$ Gyr, suppressing
the free energy driving the MTI.  However, there are observational
studies that suggest that most clusters have temperature
profiles that decline with radius outside a few hundred kpc  \citep{pratt07}. 
ÿMotivated by these results, we revisit 3D MHD simulations of
the MTI in the ICM; in contrast to previous work, we now prescribe a
fixed temperature gradient in the outskirts
of clusters.  In analyzing the results of these simulations, we focus
on quantifying the non-thermal pressure support that results from
MTI-driven turbulence.  In Section 2, we describe our methodology and
fiducial cluster model.  In Section 3, we describe the results of our
numerical experiments; we discuss their implications for SZ and X-ray
studies of galaxy clusters in Section 4.

\section{Method and Models}
We solve the usual equations of magnetohydrodynamics (MHD) with the
addition of anisotropic thermal conduction. The MHD equations in
conservative form are
\begin{equation}
\frac{\partial \rho}{\partial t} + \boldsymbol{\nabla}\cdot\left(\rho \boldsymbol{ v}\right) = 0,
\label{eqn:MHD_continuity}
\end{equation}
\begin{equation}
\frac{\partial(\rho\boldsymbol{v})}{\partial t} + \boldsymbol{\nabla}\cdot\left[\rho\boldsymbol{vv}+\left(p+\frac{B^{2}}{8\pi}\right)\mathbf{I} -\frac{\boldsymbol{BB}}{4\pi}\right] + \rho\boldsymbol{g}=0,
\label{eqn:MHD_momentum}
\end{equation}
\begin{equation}
\frac{\partial E}{\partial t} + \boldsymbol{\nabla}\cdot\left[\boldsymbol{v}\left(E+p+\frac{B^{2}}{8\pi}\right) - \frac{\boldsymbol{B}\left(\boldsymbol{B}\cdot\boldsymbol{v}\right)}{4\pi}\right] 
+\boldsymbol{\nabla}\cdot\boldsymbol{Q} +\rho\boldsymbol{g}\cdot\boldsymbol{v}=0,
\label{eqn:MHD_energy}
\end{equation}
\begin{equation}
\frac{\partial\boldsymbol{B}}{\partial t} + \boldsymbol{\nabla}\times\left(\boldsymbol{v}\times\boldsymbol{B}\right)=0,
\label{eqn:MHD_induction}
\end{equation}
where the symbols have their usual meaning. The total energy $E$ is given by
\begin{equation}
E=\epsilon+\rho\frac{\boldsymbol{v}\cdot\boldsymbol{v}}{2} + \frac{\boldsymbol{B}\cdot\boldsymbol{B}}{8\pi},
\label{eqn:MHD_Edef}
\end{equation}
where $\epsilon=p/(\gamma-1)$.  Throughout this paper, we assume
$\gamma=5/3$.  The anisotropic electron heat flux is given by
\begin{equation}
\boldsymbol{Q} = - \kappa_{\textrm{Sp}} \boldsymbol{\hat{b}\hat{b}}\cdot\boldsymbol{\nabla}T,
\label{eqn:coulombic}
\end{equation}
where $\kappa_{\textrm{Sp}}$ is the Spitzer conductivity \citep{spitz62} and
$\boldsymbol{\hat{b}}$ is a unit vector in the direction of the
magnetic field.

We simulate the fully 3D time-dependent evolution of our model galaxy
clusters using the MHD code Athena \citep{sg08} with the addition of a
module for anisotropic thermal conduction along magnetic field lines
\citep{ps05, sh07}.  Our initial condition is a spherically-symmetric,
hot, massive cluster that roughly resembles Abell 1576 (although not
fit to the exact parameters) with a mass of $1.6\times 10^{15}$ \Mo.
We use a softened NFW profile with a scale radius of $r_s = 600$ kpc
and a softening radius of 70 kpc.  We initialize an atmosphere in
hydrostatic equilibrium using the entropy power law in the ACCEPT
database for Abell 1576: a central entropy $K_0 =
186\;\textrm{keV}\,\textrm{cm}^{2}$, $K_1 =
98\;\textrm{keV}\,\textrm{cm}^{2}$, and power-law exponent, $\alpha =
1.38$ \citep{cav09}.  We use a mean molecular weight $\mu \sim 0.62$
which corresponds to a metallicity of approximately 1/3 solar.  If we
assume that our fiducial cluster is located at $z=0.1$, then for the
appropriate WMAP5 cosmological parameters $r_{500} =
1.09\;\textrm{Mpc}$, and the virial radius is $r_{200} = 1.6$ Mpc,
where $r_\Delta$ corresponds to an overdensity of $\Delta$ times the
critical density.  We do not include cooling, as we focus on the
portion of the cluster that is well outside the cooling radius.

The simulations are carried out on a $(196)^3$ Cartesian grid in a
computational domain that extends from the center of the cluster out
to $\pm 1300$ (2400) kpc for our fiducial (or larger domain)
simulations.  Within this Cartesian domain, we define a spherical
subvolume with a radius of 1225 (2200) kpc from which we extract
cluster properties, thus avoiding boundary condition effects.  In this
volume we initialize tangled magnetic fields with $\la|B|\ra =
10^{-8}\;\textrm{G}$ (plasma $\beta \sim 10^4$--$10^6$) and a Kolmogorov power spectrum (see
\citet{pqs09} for details).  We have also performed runs at $(96)^3$ and $(288)^3$ and find that the kinetic energies are reasonably well converged at our fiducial resolution.  The magnetic field amplification is not quite as well converged, and higher resolution simulations have higher final magnetic field strengths.  We choose a smaller magnetic field
strength than that observed in $z=0$ clusters as a guess of what the
magnetic field strength was when clusters first formed.  In order to
simulate clusters with a negative radial temperature gradient, we fix
the temperature at the peak of the cluster temperature profile
(approximately 200 kpc) and at the maximum radius of our model cluster
to the initially-computed temperature values.  Thus, we are imposing
Dirichlet boundary conditions on the temperature profile.  
This fixed temperature gradient then drives the continued evolution of
the MTI.  Our fiducial model has a peak temperature of 10 keV at 200 kpcs and a temperature gradient of 3.5 keV (i.e., $\Delta T/T\sim 1/3$)
over $\sim\! 1$ Mpc.  This fixed temperature boundary condition represents the
key difference between this work and \citet{psl08}.

We also examine the interplay between other sources of turbulence and the MTI.  Such turbulence could arise from galaxy wakes or structure formation, and is almost certainly present in the outer regions of clusters \citep{nagai07}.  We drive turbulence with an outer scale of 200 kpc and a steep spectrum ($v_k \sim k^{-3}$) with an energy injection rate per unit volume such that the turbulence has a Mach number, $\mathcal{M} \sim 0.13$, within 1 Mpc (more specifically $\dot{e} = 1.25\times 10^{-30} \power$).    This turbulence is driven in Fourier space with random phases with a radially constant energy injection rate using the method detailed in \citet{pqs10}.  Our toy model for turbulence results in Mach numbers somewhat smaller than found in non-conducting structure formation calculations \citep[e.g,][]{lau09}.
\section{Results}
\begin{figure}
\centering
\includegraphics[clip=true, scale=0.4]{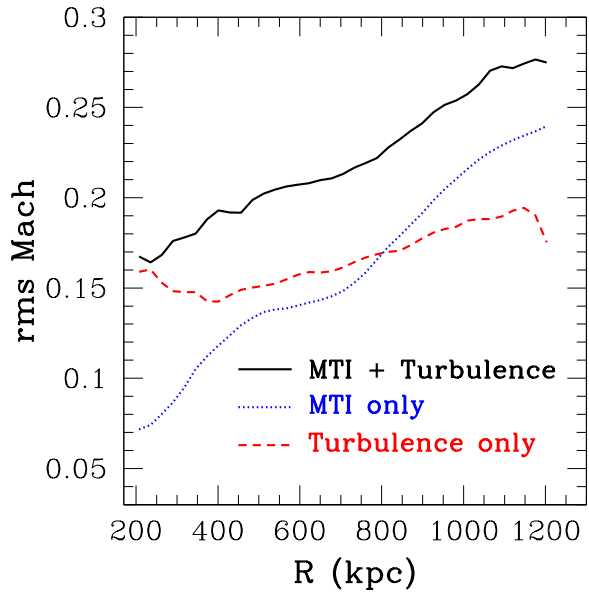}
\caption{Azimuthally-averaged rms Mach number profiles for our
  fiducial cluster model ($\Delta T = 3.5\;\textrm{keV}$,
  $r_{500} = 1.1$ Mpc).  The blue dotted line shows a simulations with
  anisotropic thermal conduction and the MTI.  The red dashed lines
  shows the same cluster model with no conduction (and no MTI), but
  with externally-driven turbulence.  The black solid line shows
  simulations with both anisotropic conduction and externally-driven
  turbulence; the turbulence resulting from the MTI and the driven
  turbulence add approximately linearly in energy (quadrature in
  $\mathcal{M}$).} \label{fig:machcomp}
\end{figure}
Our understanding of the saturation of the MTI was elucidated in
\citet{mpsq11}.  The MTI is most unstable for dynamically weak ($\beta
\gg 1$) magnetic fields that are perpendicular to gravity.  As the
instability develops, the magnetic field initially becomes
progressively more aligned with the local gravitational field.  The
initial intuition was that the MTI behaved as a dual to the
heat-flux-driven buoyancy instability \citep[HBI;][]{quat08, pq08}, and
saturated by turning horizontal ($\perp \boldsymbol{g}$) magnetic
field lines into vertical ($\parallel \boldsymbol{g}$) field lines, a
bulk reorientation of the magnetic field geometry.  However, if the
field lines are predominantly vertical, horizontal ($\perp
\boldsymbol{g}$) motions have essentially no restoring force (for weak
magnetic fields), and any small perturbation produces a horizontal
field geometry that is again MTI unstable.\footnote{Buoyant
  overstabilities, such as those highlighted in \citet{br10}, can also
  be present; however, their effect is subdominant as they grow much
  more slowly than the MTI. } Thus, the MTI cannot be suppressed by
simply rearranging the magnetic field.  Instead, in Cartesian
simulations of approximately one scale height in size, the MTI drives
sustained convection with rms Mach numbers of $\sim \!0.3$.

To understand the saturation of the MTI in global clusters with a
fixed temperature gradient, we first consider our fiducial cluster
model with the domain size of 1225 kpc.  The MTI growth time is given
by
\begin{equation}
t_{\textrm{MTI}} = \left(g \frac{\dif \ln T}{\dif \,r}\right)^{-1/2} \sim 600 \;\textrm{Myr}.
\end{equation}
If we assume that major mergers occur on a 5 Gyr timescale then there
are over 8 $e$-folding times available for the MTI to grow; thus there
is sufficient time for the MTI to become highly nonlinear.  In
approximately 2--3 Gyr, the turbulence reaches a statistical steady
state.  Figure \ref{fig:machcomp} shows an azimuthally-averaged radial
profile of the Mach number in this steady state for our fiducial
cluster model. The MTI drives large turbulent velocities with mean
Mach numbers $\simeq 0.1-0.2$ within $r_{500}$.  We also simulated our fiducial cluster with anisotropic Braginskii viscosity and found that 
the turbulent velocities are the same to within $\sim\! 5\%$.  This negligible impact of viscosity on the evolution of the MTI is consistent with theoretical predictions \citep{kunz11}.

It is also important to assess how the MTI interacts with other
sources of turbulence---turbulence that could be driven by structure
formation or other processes. 
Figure \ref{fig:machcomp} shows the saturated Mach number profiles for
numerical experiments in which we add turbulent external driving to
our fiducial model both with and without anisotropic conduction.  We
find that the externally-driven turbulence adds to the MTI-driven
turbulence such that the turbulent energies roughly add linearly.  This property corresponds to turbulent
velocities and Mach numbers that add in quadrature:
\begin{equation}
\mathcal{M}_{\textrm{tot}} \approx \left(\mathcal{M}_{\textrm{MTI}}^2 + \mathcal{M}_{\textrm{turb}}^2\right)^{1/2}.
\label{eqn:machsum}
\end{equation}
We find that this scaling holds for levels of external turbulent
driving both larger and smaller than shown in Figure
\ref{fig:machcomp}, provided that the Mach numbers are reasonably
subsonic.
\begin{figure}
\centering
\includegraphics[clip=true, scale=0.4]{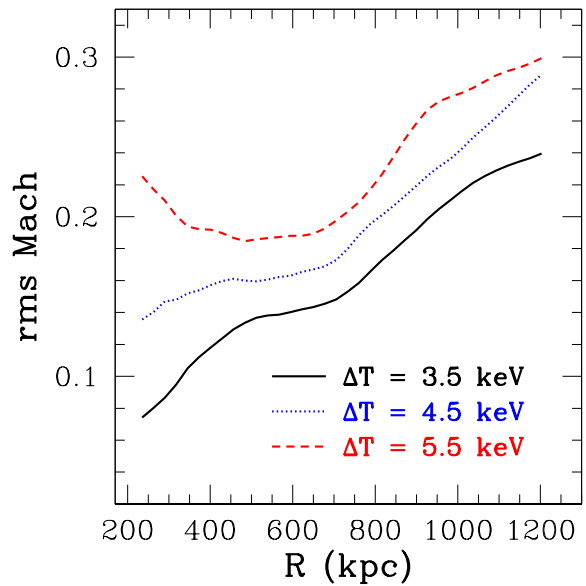}
\caption{Azimuthally-averaged rms Mach number profiles for our
  fiducial cluster model with different magnitudes of the temperature
  gradient.  The temperature gradients are labeled by the temperature
  drop from the temperature maximum at $\sim 200$ kpc to $r_{500} \simeq 1.1$ Mpc.
  The turbulent velocities scale roughly as $v\sim (\Delta T)^{1/2}$.} \label{fig:tempcomp}
\end{figure}

Although the kinetic energy quickly saturates to a relatively steady
state value for both the MTI-only and the MTI plus turbulence
simulations, the growth of the magnetic field is comparatively slow.
More quantitatively, the logarithmic growth rate of magnetic energy
for our fiducial case with both conduction and driven turbulence is
$\dif \ln \la B^2 \ra/\dif t \sim 0.4\;\textrm{Gyr}^{-1}$.  This is
$\sim 4$ times slower than the linear growth of the MTI.  Initially
the magnetic fields are completely tangled; thus, the orientation is
statistically isotropic: $\bdotr=0.5$.  For the MTI-only run, the
magnetic geometry develops a small radial bias, peaking around
$\bdotr\sim 0.65$ at 2 Gyr before declining to isotropy again.  For
the run with the MTI and driven turbulence, the magnetic field only
briefly deviates from statistical isotropy, fluctuating around $\bdotr
= 0.5$.

Previously \citet{rusz11} performed a cosmological study of the
evolution of a single cluster with anisotropic thermal conduction.
They found evidence for a small radial bias in the magnetic field;
however, in the cosmological simulations the effects of the MTI and
radial infall were not disentangled.  In light of our simulations,
it is likely that the radial magnetic fields in the cosmological
simulation are due to infalling substructure dragging magnetic field
lines out.  The isolated non-cosmological simulations presented here
are necessary to separately understand these effects.  Our results
also suggest that the radial magnetic field bias observed in Virgo
\citep{pd10} is also likely due to infall or galaxy motions rather
than the MTI. 

For a given cluster mass, the saturated state of the MTI depends
primarily on $\dif\ln T/\dif \ln r$, since the free energy of the
temperature gradient drives the MTI.  We adjust our initial
equilibrium to several different temperature gradients by increasing
the temperature maximum to study this effect, holding all of the
remaining model parameters fixed.  Figure \ref{fig:tempcomp} shows
that there is a clear trend in the resulting turbulence driven by the
MTI, with larger temperature gradients producing stronger turbulence.
If we calculate a power-law fit to the turbulence induced by the MTI
at $r_{500}$ we find that $\mathcal{M} \propto \left(\Delta T\right) ^{0.5}$,
where $\Delta T$ is the temperature drop across 1 Mpc.  
This scaling is consistent with the simple mixing length
estimate of $v^2 \propto \Delta T$.  

\begin{figure}
\centering
\includegraphics[clip=true, scale=0.4]{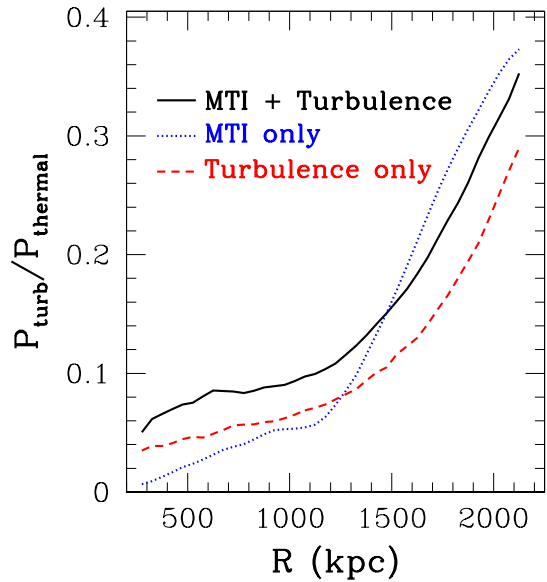}
\caption{Azimuthally-averaged, steady-state turbulent pressure
  profiles (normalized to the thermal pressure) for the larger volume
  simulations of our fiducial cluster model ($\Delta T =
  3.5\;\textrm{keV}$, $r_{500} = 1.1$ Mpc).  The legend is
  the same as Figure \ref{fig:machcomp}.  At $r \gtrsim
  r_{500}$, the turbulent pressure support can be a significant
  fraction of the thermal pressure.} \label{fig:bigbox}
\end{figure}

Much of the contribution of clusters to the SZ power spectrum comes
from scales comparable to or larger than the virial radius of groups
and clusters.  In order to understand the possible effects of the MTI
on such SZ power spectrum measurements, we have extended our
simulation domain to a larger size ($\sim \!2 \,r_{500}$).  In these
larger domain simulations, the MTI drives even more vigorous
convection.  Figure \ref{fig:bigbox} shows the steady-state turbulent
pressure profiles for our large domain simulations with the MTI alone,
externally-driven turbulence alone, and both together.  The MTI is
able to drive convection that yields significant non-thermal pressure
support, reaching $\sim$ 35\% of the thermal pressure near $2\,
r_{500}$.  The turbulent Mach numbers increase significantly towards
the cluster outskirts.  Our interpretation is that large convective
motions deep in the core overshoot and continue with large momentum to
the low-density cluster outskirts; this effect was also seen in our earlier stratified Cartesian simulations \citep{mpsq11}.

More quantitatively, in a shell centered at 1800 kpc of width 50 kpc,
we find that the rms Mach number with 1$\sigma$ fluctuations is $\la
\mathcal{M} \ra \approx 0.54 \pm 0.19$.  The maximum Mach number in
this shell reaches 1.26.  The behavior of the simulation with both the
MTI and turbulence simultaneously is particularly interesting---within
approximately 1200 kpc, the MTI-driven turbulence and external
turbulence add in quadrature as seen previously in Figure
\ref{fig:machcomp}; however, beyond this radius the net turbulence is
approaching trans-sonic and the MTI is unable to grow effectively.
If we perform the same experiment with the higher temperature gradient
of $\Delta T = 5.5$ keV, we find that the Mach number can reach values
as high as $\mathcal{M}\sim 2.5$, also likely due to overshoot.
Unfortunately, we are not able to run these simulations to a full
steady-state as the MHD integrator in Athena crashes in high Mach
number, highly stratified, turbulence.
\section{Discussion and Implications}

Galaxy clusters have the potential to serve as powerful probes of
structure formation by breaking the $\sigma_8$--$\Omega_M$ degeneracy
and constraining the dark energy equation of state, $w$.  Moreover,
much of the constraining power of galaxy clusters is highly
complementary to constraints from the CMB and other methods.  However,
for both SZ and X-ray observations, these applications require
accurately determining the total cluster mass either directly through
reconstructing a hydrostatic mass or indirectly through calibrated
scaling relations (note that this is, of course, not true for lensing
mass determinations).  It is critical to understand any systematic
errors or biases in these cluster mass determinations.

Cosmological hydrodynamic simulations have shown for some time that
turbulence produced by structure formation can provide significant
non-thermal pressure support in clusters, particularly at large radii
\citep[e.g.,][]{evrard90,vazza09}.  Our results demonstrate that the
MTI, a convective instability triggered by
anisotropic thermal conduction along magnetic field lines, can also
produce significant (near-sonic) turbulent motions in the outer parts
($\gtrsim 200$ kpc) of clusters. This turbulence is the strongest at
large radii $\sim r_{200}$ (Fig. \ref{fig:bigbox}), in part because of
convective overshooting from smaller radii.  The non-thermal pressure
support produced by the MTI is not directly accounted for in current attempts
to estimate cluster masses from observables such as the X-ray
luminosity or SZ signal.

Mass measurements of galaxy clusters using \textit{Chandra} or
\textit{XMM-Newton} have typically been confined to within $r_{500}$
where the x-ray surface brightness is reasonably large.  Even within
this radius, we find non-neglible turbulent pressure support $\sim
\!3$--10\% (Fig.~\ref{fig:tempcomp}) due to the MTI, with the exact
value depending on the cluster temperature gradient.  X-ray
observations of clusters have been used to try to constrain the
properties of dark energy (e.g., \citealt{allen08}); our results
suggest the need for somewhat larger priors on the fraction of
non-thermal ICM pressure support in such modeling.  

The thermal SZ effect is often characterized by the Compton-$y$
parameter, which measures the fractional change of the CMB temperature
and is $\propto n_e T_e$.
The total SZ signal is then simply proportional to the total thermal
pressure of the cluster plasma.  As a result, the SZ signal correlates
well with the total cluster mass in cosmological hydrodynamics
simulations (e.g., \citealt{nagai06}).  Turbulence due to the MTI
would decrease the SZ signal for a given cluster mass relative to
these scaling relations; it would also likely introduce additional
scatter into such relations.

In addition to finding individual clusters, SZ experiments survey the
entire sky to produce an angular power spectrum whose amplitude is
very sensitive to $\sigma_8$, scaling as $C_{\ell} \propto \sigma_8^7
(\Omega_b h )^2 $ \citep{ks02}.  The connection between $\sigma_8$ and
the SZ power spectrum requires calibration by simulations, and the
first SZ results 
found $\sigma_8=0.746 \pm 0.017$, a value that was statistically in
tension with other probes of $\sigma_8$ \citep{lueker10, sehgal11}.
Resolving this discrepancy requires decreasing the predicted SZ power
by a factor of $\sim \!\!2$.  Both the ACT and SPT groups have attempted
to remedy this model-specific deficiency by running cosmological
simulations with additional feedback \citep{shaw10, battaglia10}.
Analytical modeling by \citet{shaw10} shows that increasing the
fraction of non-thermal pressure support in galaxy clusters can
significantly reduce SZ power at large angular scales.  Thus, the
additional non-thermal pressure support produced by the MTI at large
radii in galaxy clusters (Fig. \ref{fig:bigbox}) may significantly
modify the determination of $\sigma_8$ from the SZ power spectrum.

In Figure \ref{fig:tempcomp} we have shown that there is a direct
connection between the level of turbulent non-thermal pressure support
due to the MTI and the cluster temperature gradient at large radii.
This is a simple consequence of the fact that it is the temperature
gradient that drives the instability in the first place.  This result
is particularly interesting because it implies that constraints on the
temperature gradient could be used to estimate the MTI contribution to
the turbulent pressure, which would be very valuable in improving
cluster mass estimates.  Absent such observational constraints,
however, Figure \ref{fig:tempcomp} implies that variations in cluster
temperature gradients as a function of mass and/or redshift could
introduce a subtle bias in the X-ray and SZ signals used to constrain
cosmological parameters.  In addition, in lower mass clusters or groups with lower virial temperatures, the thermal conduction time
across the cluster outskirts will be longer than in the massive cluster model considered in this work.   This is likely to cause the
turbulent pressure support induced by the MTI to depend explicitly on cluster mass.   Recent X-ray observations of Perseus with the \textit{Suzaku} telescope have shown evidence that suggests clumping of baryons near the virial radius \citep{simionescu11}.  Turbulence can drive fractional density enhancements ($\delta \rho/\rho)$ that scale as either $\mathcal{M}^2$ for compressive fluctuations or linearly as $\mathcal{M}$ if the fluctuations are behaving as a passive scalar in the stratified atmosphere.  While not large enough in magnitude alone, the turbulence we find can contribute to these density enhancements.

In our calculations we have fixed the temperature gradient at large
radii in our model clusters by using Dirichlet boundary conditions.
In reality, the temperature gradient is likely set by an interplay
between structure formation (e.g., the virial shock) and thermal
conduction itself.  In previous simulations of isolated clusters with
anisotropic thermal conduction that did not fix the temperature
gradient, thermal conduction was effective enough to wipe out the
temperature gradient after $\sim \!1$ Gyr, strongly suppressing the
level of turbulence generated by the MTI \citep{psl08}.  The
observational evidence for non-zero temperature gradients at large
radii in clusters \citep[e.g.,][]{pratt07} motivated our choice of
boundary conditions in this work.  Physically, we regard this choice
as a proxy for cosmological physics not included in our simulations
(e.g., heating by substructure moving through the ICM and/or
compressional heating and inflow after the virial shock).  It remains
to be seen, however, whether significant temperature gradients can
indeed be maintained.  The only cosmological simulation with anisotropic
thermal conduction to date found $\dif T/\dif r < 0$ at large
radii, with the magnitude of the temperature gradient comparable to
that assumed in this work (see Fig. 2 of \citealt{rusz11}).

The non-cosmological simulations presented in this paper are not
sufficient to accurately determine the contribution of the MTI to the
turbulent pressure support in clusters without calibration by comparison to cosmological
simulations.  This is particularly true at large radii $\sim\! r_{200}$
where our isolated cluster simulations predict near sonic motions for
even modest temperature gradients (Fig. \ref{fig:bigbox}).  We suspect
that the convective overshooting that produces these large velocities
at large radii is likely to be generic.  In the future, a suite of
cosmological simulations with anisotropic thermal conduction will be
necessary to fully assess the effects of the MTI on SZ and X-ray
observations of clusters.  Such simulations are now possible but isolated cluster simulations like those presented
here are still necessary to pinpoint and understand the key physics.


\acknowledgments We thank Daisuke Nagai, Marcus Br\"{u}ggen, and
Christoph Pfrommer for useful conversations that helped steer this
work.  Support was provided in part by NASA Grant ATP09-0125, NSF-DOE
Grant PHY-0812811, and by the David and Lucille Parker Foundation.
Support for P.S. was provided by NASA through the Chandra Postdoctoral
Fellowship grant PF8-90054 awarded by the Chandra X-Ray Center, which
is operated by the Smithsonian Astrophysical Observatory for NASA
under contract NAS8-03060.  We would like to thank the hospitality of
the KITP where much of this work was performed and supported in part
by the National Science Foundation under grant PHY05-51164.  The
computations for this paper were perfomed on the \textit{Henyey}
cluster at UC Berkeley, supported by NSF grant AST-0905801 and through
computational time provided by the National Science Foundation through
the Teragrid resources located at the National Institute for
Computational Sciences under grant TG-AST080049.

\bibliography{cluster}
\end{document}